# Parity Fluctuations Between Coulomb Blockaded Superconducting Islands

Boldizsár Jankó and Vinay Ambegaokar
*Laboratory of Atomic and Solid State Physics, Cornell University, Ithaca, New York 14850-2501*
(May 8, 1995)

We find that if two superconducting islands of different number parity are linked by a tunnel junction the unpaired electron in the odd island has a tendency to tunnel into the even island. This process leads to fluctuations in time of the number parity of each island, giving rise to a random telegraph noise spectrum with a characteristic frequency that has an unusual temperature dependence. This new phenomenon should be observable in a Cooper-pair pump and similar single-electron tunneling devices.

In this paper we study the problem of tunneling between number restricted superconducting islands. Small capacitatively isolated superconducting islands have recently attracted much experimental [1] and theoretical [2,3] attention. The basic question addressed in various ways by these studies is: how are the properties of a superconducting sample affected by the evenness or oddness of the number of electrons it contains? Since the Bardeen-Cooper-Schrieffer (BCS) [4] ground state is made from a coherent superposition of Cooper *pairs* there will always be an unpaired particle present if the total number is odd. In the experiments cited the effects of number parity are revealed by using the Coulomb energy to control the addition or removal of a single electron.

When an unpaired particle is present, the energy difference between an odd and an even sample at absolute zero will always be larger by an amount $\Delta$ than it is in the normal state: $E(2N+1) - E(2N) - \mu = \Delta$ no matter how big N is (here $\Delta$ and $\mu$ are the superconducting energy gap and chemical potential, respectively). At low, but finite temperatures, the free energy difference between the two samples is equal to $\delta \mathcal{F}_{odd-even} = \Delta - k_B T \ln N_{eff}$. The entropy term $-k_B T \ln N_{eff}$ arises because the unpaired particle can sample the available states (approximately $N_{eff}$ in number). It is only a logarithmic function of N since $N_{eff} \simeq 2N(0)\mathcal{V}\sqrt{2\pi T\Delta} \propto N$ [1–3]. [Here $N(0)$ is the single particle density of states at the Fermi level, and $\mathcal{V}$ is the volume of the island.] At $T^* = \Delta/\ln N_{eff} \sim 200 - 300 mK$ the free energy difference practically vanishes (for typical island parameters see Ref. [1]). As a result parity effects are robust against size; they have been observed [1] in the equilibrium and transport properties of islands with particle numbers as large as roughly $N \sim 10^9$. The constraint on N comes from the Coulomb blockade condition of thermal fluctuations $k_B T \ll e^2/2C$.

Now consider two islands between which electrons can tunnel. The particle number and parity of each individual island is no longer fixed, although the total number and parity of the system is conserved. The situation greatly simplifies at low temperatures, when very few quasiparticles are present. Cooper pair tunneling, which does not affect the parity of the island, corresponds to a twice as large charge fluctuation and can be suppressed by the Coulomb blockade even when the charging energy of the system is tuned by gate voltages to permit single particle tunneling. The most interesting case is when initially one of the islands is in the even, and the other one in the odd parity state. When $T < T^*$, all the particles are paired up on the even island, and on the average there is only one quasiparticle present on the odd island. If the tunnel resistance $R_T \gg R_Q = h/e^2$ [5], the unpaired electron is well localized on the odd island. Nevertheless, there is a nonzero transition rate for tunneling of a single quasiparticle, always from the odd island to the even island. This leads to a random telegraph noise type fluctuation in time of the number parity, with a Lorentzian noise spectrum [6].

An important remark should be made at this stage. This parity oscillation in this regime is not a coherent quantum oscillation: the state of the unpaired particle is not given by a phase coherent superposition of its localized wave functions on the left and right islands. Because there are many available states for the particle on each island the transition rate must be evaluated by Fermi's Golden Rule and no phase coherence can be preserved under these circumstances. This incoherent tunneling process, however, as we will show later on, has some very peculiar properties: it sets in below $T^*$ and the tunneling rate either has a maximum or saturates as the temperature is lowered, depending on how strong the coupling is between the system and the environment. Such a behaviour is quite unusual and cannot be easily sorted into the any of the usual classes, like that of thermally activated processes [7].

We will suggest experimental setups in which this spectrum could be measured. Our analysis gives new stability regions for the Cooper pair pump [8] and other devices affected by parity effects and implies modified parameters for optimal operation. The coupling between the tunneling electron and the collective modes of the electromagnetic environment turns out to be very important. Our calculations show that when this coupling is too strong the unpaired electron becomes trapped on one island.

Our model consists of two superconductors of different parity, even and odd, linked by a tunnel junction. For the sake of simplicity, we take these superconductors to be otherwise identical: the same volume $\mathcal{V}$, energy gap $\Delta$ etc. We want to describe the tunneling events taking place in this junction. Since electrons have charge, a tunneling event perturbs the equilibrium charge distribution, and excites environmental modes: oscillations of the electromagnetic field in the circuit (or electromagnetic environment) connected to the junction. Tunneling in the presence of environment is assumed to be adequately described by the Hamiltonian [9,10]:

$$H_T = H_{T,l\to r} + H_{T,r\to l}$$
$$= \sum_{k,q,\sigma} T_{kq} c^\dagger_{q\sigma} c_{k\sigma} \Lambda_e + T^*_{kq} c^\dagger_{k\sigma} c_{q\sigma} \Lambda^\dagger_e. \quad (1)$$

The first term describes the tunneling with amplitude $T_{kq}$ of an electron in state $k\sigma$ from the left into $q\sigma$ on the right. The second terms corresponds to the the time-reversed process. We assume zero magnetic field and make the arbitrary choice that the superconductor on the left is odd and the one on the right is even in particle number. $\Lambda_e$ is a charge displacement operator which satisfies $\Lambda_e Q \Lambda^\dagger_e = Q - e$. The quantity of interest is the transition rate of the system from this initial state into one in which the parity of the superconductors is switched. The total rate is the difference of quasiparticle tunneling rate from left to right, $\Gamma^\to(V)$, and from right to left, $\Gamma^\leftarrow(V)$. At low temperatures and zero bias voltage this rate, in contrast with the usual situation, is nonzero and is directed from the odd superconductor towards the even one (in this case from left to right):

$$\Gamma \equiv \Gamma^\to(0) - \Gamma^\leftarrow(0) = \frac{1}{2e^2 R_T N_{eff}} \int_{-\infty}^{+\infty} dE dE'$$
$$\times \frac{N_S(E) N_S(E')}{N(0)^2} (f_o - f_e) P(E - E'), \quad (2)$$

where $R_T^{-1} = 4\pi(e^2/h)|T|^2$, and $R_T$ is the tunneling resistance. It is assumed that $|T_{kq}|^2$ is a weak function of momentum and can be replaced by its average $|T|^2$ around the Fermi energy. The quasiparticle density of states is given by the BCS result [4]

$$\frac{N_S(E)}{N(0)} = \frac{|E|}{\sqrt{E^2 - \Delta^2}} \text{ for } |E| > \Delta, \quad (3)$$

and zero otherwise. The function $P(E)$ is the Fourier transform of $\langle \Lambda^\dagger(t) \Lambda(0) \rangle$ [10]. In physical language, $P(E)dE$ gives the probability that the electron exchanges an energy in the range $E, E + dE$ with the environment during the tunneling process.

A similar process occurs in the ordinary quasiparticle tunneling [11] when thermally excited quasiparticles tunnel between macroscopic superconductors giving rise to a structure in the I-V curve at $V = \pm|\Delta_1 - \Delta_2|$. In this case there are no thermally excited quasiparticles but there is one available purely due to the parity constraint. The difference in the distribution functions $f_o - f_e$ which enters the above expression, is [3]

$$f_o - f_e = \frac{\text{cosech}\beta E_k}{\prod_{k\sigma} \coth\frac{\beta E_k}{2} - \prod_{k\sigma} \tanh\frac{\beta E_k}{2}}. \quad (4)$$

At low temperatures the difference in quasiparticle occupation numbers becomes $f_o - f_e = \exp[-\beta(E_k - \Delta)]/2N_{eff}$ whereas close to $T^*$ the denominator in (4) becomes very large and the parity difference becomes negligible.

While it is not possible to compute $P(E)$ in a general case, there are two important and physically intuitive limits to consider. For low environmental impedance $|Z(\omega)| \ll h/e^2$ the dominant process is elastic tunneling: $P(E - E') = \delta(E - E')$. The environmental charge configuration remains the same, similar to the recoilless Mössbauer transition. If, however, $|Z(\omega)| \gg h/e^2$, at low temperatures $T \ll E_c$, $P(E)$ becomes

$$P(E) = \delta(E - E_c), \quad (5)$$

where $E_c$ is a charging energy scale (note that an interpolation between the two limits is possible formally turning $E_c$ to zero). In the latter limit for low temperatures $T \ll T^* \ll E_c$ we obtain

$$\Gamma = \frac{\Delta \exp(-\beta E_c)}{2e^2 R_T N_{eff}}(1 + e^{-\beta\Delta a}) \times$$
$$\int_1^{+\infty} \frac{(x+a)xe^{-\beta\Delta(x-1)}}{\sqrt{(x^2-1)[(x+a)^2-1]}}, \quad (6)$$

where $a \equiv E_c/\Delta$. When $a = 0$ the integral diverges logarithmically, as does the ordinary quasiparticle current between two macroscopic superconductors [11] due to the overlapping BCS density of states. It is quite well understood for a long time [12] that in a real metal the divergence is naturally removed by gap anisotropy and nonzero quasiparticle lifetime. For nonzero $a$ the rate is further suppressed by the overal factor $\exp(-\beta E_c)$. The odd electron is "locked" by the electromagnetic environment into one island: a tunneling event in the presence of a high impedance environment would result in the creation of a large number of low energy environmental modes so that the initial environmental state is practically orthogonal to the required final state. As a consequence, the transition amplitude is decreased by this "orthogonality catastrophe" [13].

The dependence of the fluctuation rate on other parameters is clear from the above result and easy to interpret. The process is slowed down by increasing tunnel resistance, which corresponds to decreasing the transparency of the junction. The time spent by the odd particle in one superconductor, $\Gamma^{-1}$ is proportional to the volume of the superconductor via $N_{eff}$, as it should be. In order for the process to be observable, the temperature has to be low enough, $T < T^*$, so that parity effects are important. Below this temperature the rate is proportional to the energy gap.

The complete temperature dependence of $\Gamma(T)$ is shown in figure 1, calculated from eqs. (2), (4) and (5). First note that the rate becomes nonzero only below $T^*$ and has a maximum before it becomes exponentially suppressed according to eq.(6) due to the coupling to the environment. The parameters were chosen from reference [8] for later convenience, when we discuss these ideas in the context of the Cooper pair pump. If the coupling is weak (see inset), the rate levels off to a very high value limited by gap anisotropy or impurity scattering. These effects smear out the square-root singularity in the BCS density of states. We treat this limit by formally taking the parameter $a$ to a very small but nonzero value. Such temperature dependence is quite unusual for incoherent quantum tunneling process [14] where the exponential dependence (corresponding to thermally activated hopping) at higher temperatures is taken over by power law dependence $\Gamma \propto T^{2\alpha-1}$ at low temperature.

Let us now turn to the discussion of the single-Cooper-pair pump [8]. In our opinion, it has all the necessary ingredients to exhibit parity fluctuations. It is the simplest device that has two distinct islands (see Fig. 2), and can be charged in a locally stable manner - by appropriately tuning the gate voltages $U_1$ and $U_2$ - to a configuration characterized by $(n_1, n_2)$. (Here $n_1$ and $n_2$ are the number of excess electrons on each island). We have calculated the regions of stable charging in the $(U_1, U_2)$ plane. Our result is different from the previous ones [8] in that we now take into account the increase in the free energy of those configurations which unpaired electrons are present. In each such region the system has to be stable against tunneling one each junction $i$ to right (left) $i+, (i-)$ (here $i = \{1, 2, 3\}$)

$$\Delta F^{i\pm} = \Delta E^{i\pm} + \Delta[p(n_1) + p(n_2)].\mathcal{F}_{e/o} > 0, \qquad (7)$$

That is, the change in the total free energy of the system $\Delta F^{i\pm}$ equals to the sum of changes in the charging energy $\Delta E^{i\pm}$ and parity dependent energy, $\delta[p(n_1)+p(n_2)]$, where $p(n) = 1$ if $n$ is odd and $p(n) = 0$ if $n$ is even. As the temperature is lowered below $T^*$ the regions with odd island charge configuration will shrink, especially those with both $n_1$ and $n_2$ odd. The result of the calculation is shown in Fig. 3 for a particular value of the parameter $d \equiv \mathcal{F}_{e/o}/(e^2/3C_0) = 1/2$ when the even-odd boundaries (dotted lines) have a maximal length for nonzero $\mathcal{F}_{e/o}$. ($C_0$ is the capacitance of the tunnel junctions, assumed to be equal). At this stage the all-odd configurations, like $(1,1)$, disappear from the ground state configuration. We obtain a very simple condition necessary for $2e$ charge quantization: $\Delta > e^2/3C_0$. This is desirable for the successful operation of the Cooper-pair pump since then only the all-even regions survive the low temperature limit. The condition seems not to be fulfilled in ref. [8] where even at the lowest operating temperatures ($50mK$) $d \sim 0.9$.

Assuming that both islands have an even number of electrons in the configuration $(0,0)$, we are interested in the case of $n_1$ odd and $n_2$ even (or vice versa). The system must to be tuned via $U_1$ and $U_2$ at the lines separating the stability regions $(2n+1, 2n)$ from $(2n, 2n+1)$ (dotted lines in Fig 3). The circle shows a typical load curve around which the Cooper pair pump is driven. Clearly, the system is unstable against parity fluctuations. An error in the current occurs if such an event forces the system to leave the circle. Previous studies [8] of the Cooper-pair pump reported difficulties connected to the presence of quasiparticles at very low temperatures. Parity effects might be at the origin of these difficulties, since an island with odd number of particles contains at least one quasiparticle, no matter how low the temperature is.

By coupling the islands to single-electron electrometers, the process should be detectable by measuring the island charges as a function of time and extracting the noise spectrum. Above $T^*$ there should be no effect, and the values of $n_1$ and $n_2$ should evolve relatively smoothly from one value to the other one, as the system is driven accross such boundary. Assuming, as is reasonable, that there is no correlation between the parity flip events and that the probability for such a transition in the time interval $t, t+dt$ is $\Gamma(T)dt$, the parity fluctuation process will generate telegraph noise [6] with its characteristic Lorentzian noise spectrum $S(\omega) = \frac{1}{2}\Gamma(\omega^2 + 4\Gamma^2)^{-1}$. Focusing on the middle junction which controls parity fluctuation between the islands, we approximate the circuit by replacing the outer junctions with capacitors [9]. Then the charging energy scale used in eq. (5) becomes $E_c = 1/3(e^2/2C_0)$ and our result (6) for $\Gamma(T)$ can be directly used. Furthermore, we assume that the time between each tunneling event is larger than the characteristic relaxation time, so that it is justifiable to use the equilibrium environmental states in calculating the tunneling rates. We have not considered possible quasiparticle tunneling events on one of the outer junctions since they can be neglected in a system with significant Coulomb blockade. Higher order processes, like co-tunneling, as well as quantum fluctuations of the island charge are only important when the tunnel resistance is comparable to the resistance quantum.

We thank M. Devoret, L. Levitov, A. Louis, J. Mooij and M. Tinkham for helpful comments. This work was supported in part by the MRL Program of the NSF under award No. DMR-9121654.

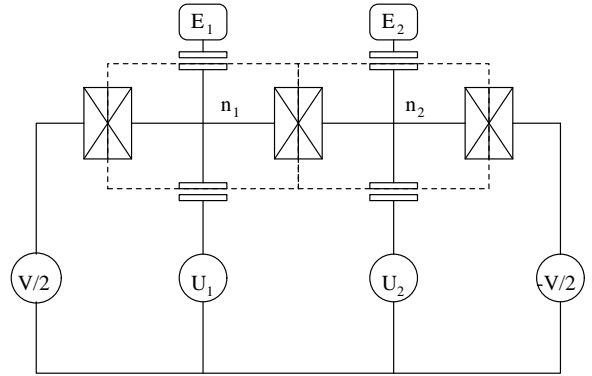

Figure 2.
FIG. 2. Schematic circuit diagram of the single Cooper pair pump, with excess electrons $n_1$ and $n_2$ on islands and delimited by dashed lines. Gate voltage sources $U_1$ and $U_2$ are coupled to the islands via two capacitors. The parity fluctuation is mediated by the middle junction, and is monitored by the electrometers $E_1$ and $E_2$. In this paper we take $V = 0$.

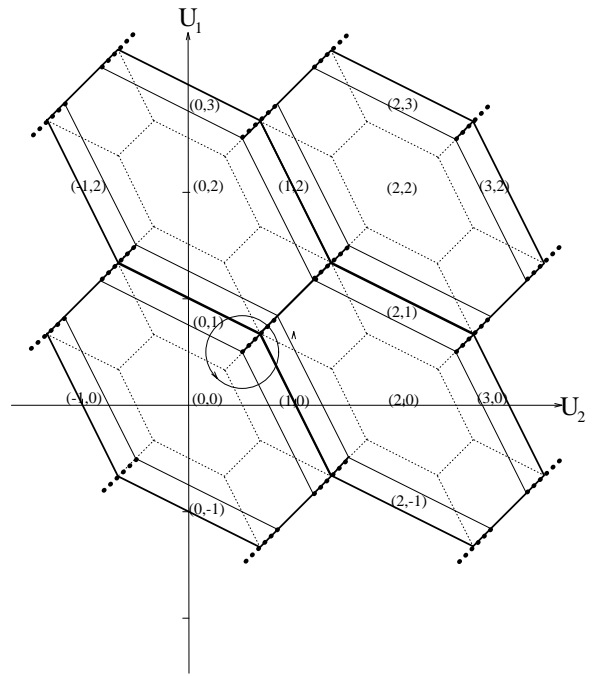

Figure 3.
FIG. 3. Stability domains for the ground state charge configuration of the single Cooper-pair pump, in the plane of gate voltages $(U_1, U_2)$. Numbers $(n_1, n_2)$ correspond to the number of excess electrons on each island. Thin dashed lines give boundaries for stable single electron charging, whereas thick solid lines give the Cooper-pair stability regions. The dotted lines mark the boundary where parity fluctuation should be observed. The circle is a typical cycle load curve used to operate the Cooper-pair pump

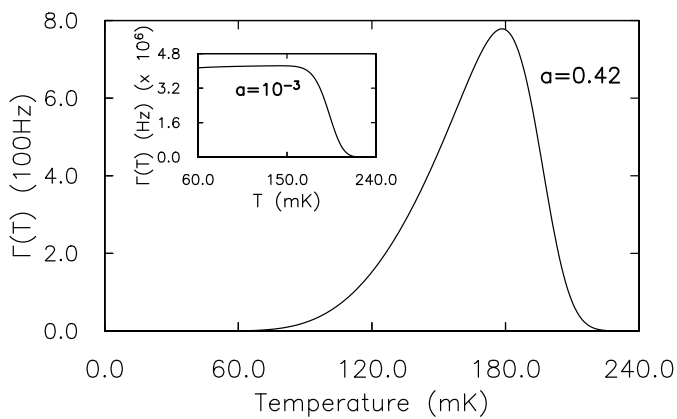

FIG. 1. The temperature dependence of the parity fluctuation rate. For high environmental impedance $|Z(\omega)| \gg h/e^2$. The inset shows the case when $|Z(\omega)| \ll h/e^2$. (parameters from Geerlings et al.: $T^* \sim 220 mK$, $N_{eff} \sim 10^4$, $R_T = 85 k\Omega$, $\Delta = 2.31 K$, $a = E_c/\Delta = 0.42$, see eq(7) and text).